\documentclass[12pt,preprint]{aastex}

\slugcomment{published in \emph{Icarus}, \textbf{188}: 35--46 (2007) | doi: 10.1016/j.icarus.2006.10.037}

\shorttitle{Measurements of the GRS with Automated Feature Tracking}
\shortauthors{Choi et al.}

\begin{document}

\title{Velocity and Vorticity Measurements of Jupiter's Great Red Spot Using
Automated Cloud Feature Tracking}


\author{David S. Choi}
\affil{Department of Planetary Sciences, University of Arizona, Tucson, AZ 85721}
\email{dchoi@lpl.arizona.edu}

\author{Don Banfield and Peter J. Gierasch}
\affil{Department of Astronomy, Cornell University, Ithaca, NY 14853}

\and

\author{Adam P. Showman}
\affil{Department of Planetary Sciences, University of Arizona, Tucson, AZ 85721}

\begin{abstract}
We have produced mosaics of the Great Red Spot (GRS) using images taken by the \emph{Galileo} spacecraft in May 2000, and have measured the winds of the GRS using an automated algorithm that does not require manual cloud tracking. Our technique yields a high-density, regular grid of wind velocity vectors that is advantageous over a limited number of scattered wind vectors that result from manual cloud tracking. The high-velocity collar of the GRS is clearly seen from our velocity vector map, and highest wind velocities are measured to be around 170 m s$^{-1}$. The high resolution of the mosaics have also enabled us to map turbulent eddies inside the chaotic central region of the GRS, similar to those mapped by \citet{Sada96} and \citet{Vasavada98}. Using the wind velocity measurements, we computed particle trajectories around the GRS as well as maps of relative and absolute vorticities. We have discovered a narrow ring of cyclonic vorticity that surrounds the main anti-cyclonic high-velocity collar. This narrow ring appears to correspond to a ring surrounding the GRS that is bright in 5-$\mu$m \citep{Terrile79}. It appears that this cyclonic ring is not a transient feature of the GRS, as we have discovered it in a re-analysis of \emph{Galileo} data taken in 1996 first analyzed by \citet{Vasavada98}. We also calculate how absolute vorticity changes as a function of latitude along a trajectory around the GRS and compare these measurements to similar ones performed by \citet{Dowling88} using \emph{Voyager} data. We show no dramatic evolution in the structure of the GRS since the \emph{Voyager} era except for additional evidence for a counterrotating GRS core, an increase in velocity in the main velocity collar, and an overall decrease in the length of the GRS.
\end{abstract}



\section{Introduction}

The \emph{Galileo} spacecraft began its mission in 1995 and orbited
Jupiter for eight years, during which it observed the planet's
satellites, rings, and atmosphere. During the 28th orbit of
\emph{Galileo} in May 2000 (G28), the spacecraft imaged Jupiter's
Great Red Spot (GRS) with a remarkable level of detail. An observation
sequence was designed to make three observations of the vortex over a
span of about two hours, and resulted in one of the highest resolution
images of the GRS ever created. We present mosaics of these imaging
observations that were used for further data analysis. A further
motivation behind our analysis was to take advantage of the high
native resolution of these mosaics ($\sim$12 km pixel$^{-1}$) and use
an automated cloud feature tracking technique in order to avoid the
errors and disadvantages associated with cloud tracking by hand. We
will describe the automated technique that we utilize and how we
optimized it to return the most accurate results possible, and how we
compensated for unavoidable blemishes in the images, such as the
inadvertent presence of a moon shadow. We present the results of our
analysis and include a discussion on techniques that we employed in
order to minimize the errors associated with our analysis. We also
include a discussion on how our observations fit in the context of
previous imaging studies of Jovian atmospheric dynamics, and what we
hope to accomplish with this technique in the future.

\section{Observations}

\subsection{Image Processing, Mosaic Construction}

Three observations of the GRS at approximately 1 hour intervals were
taken during the 28th orbit of \emph{Galileo} using the Solid State
Imaging (SSI) Camera. We removed any bad pixels present in the images
and divided each image by a flat field. All images were observed using
the near-IR (756 nm) filter onboard the SSI system. Although all of
the images are high resolution, some images were less than ideal. Some
images from the first two observations contain Europa's shadow, and a
targeting error resulted in the GRS being off-center in the images and
failing to observe the northern outer edges of the GRS. (Note:
Although \citet{Simon-Miller02} state that this shadow is Io's, we
have used the JPL Solar System Simulator (space.jpl.nasa.gov) to
determine that it is, in fact, Europa's shadow.)

\begin{figure}[htb]
  \centering
  \includegraphics[width=5.5in, keepaspectratio=true]{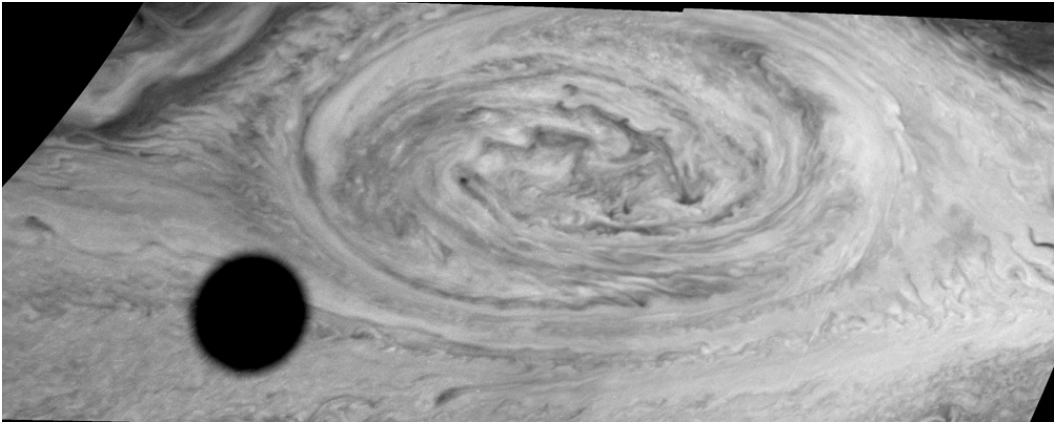}
  \caption[GRS Mosaic 1]{
    \label{Figure: Mosaic}
    One of the mosaics analyzed by our feature tracker. The shadow in
    the mosaic belongs to Europa.
    }
\end{figure}

In addition to rectifying problems with the images, we used the NAV
and NAV2 subroutines under the VICAR software package to make subtle
corrections to the navigational data that accompanied each image to
attain the highest level of accuracy possible for our data
analysis. This data consisted of spacecraft orientation and camera
pointing information such as range, subsolar point coordinates, and
position angle. We used planetary limb-fitting for one of the raw
images as the basis of the image navigation, and supplemented the
corrections to the navigation using common points found among raw
image pairs. Once all of the corrected navigational data was compiled
for each image, a mosaic for each observation was constructed using
MaRC (Map Reprojections and Conversions), an open-source map-making
software package.\footnote{Interested readers can download MaRC at
http://sourceforge.net/projects/marc} MaRC is designed to generate
mosaics in a variety of map projections using the available
navigational and camera orientation data. We built the mosaics using a
simple cylindrical projection in order to facilitate the data
analysis. Our set of three mosaics is 3000x1200 pixels and is centered
on the GRS itself. Each mosaic spans 15 to 27 degrees south
planetocentric latitude\footnote{All latitudes referenced in this
paper in the text or in the figures are planetocentric unless stated
otherwise.}, and 12 to -18 degrees west longitude (System III). The
native resolution of the mosaics is roughly 12 km pixel$^{-1}$. Once
all of the mosaics were constructed, we made one final correction to
the mosaics by dividing by the cosine of the solar incidence angle
($\mu_o$) at the time the data were taken. In an effort to minimize
manipulation to the imaging data, no further contrast enhancement was
performed on the images; tests using high-pass filtered mosaics showed
no systematic improvement in the final results. One of our mosaics is
shown in Figure~\ref{Figure: Mosaic}.

\subsection{Automated Cloud Feature Tracking}

Previous methods of extracting information on the atmospheric winds
relied primarily on manual cloud tracking, such as that performed by
\citet{Dowling88}. A human operator selects the same cloud feature in
an image pair, and software is then used to calculate the wind
velocity at that location. An automated approach removes the human
selection component. We have developed an algorithm that extracts a
basis portion of an image mosaic, numerous portions of a mosaic at a
later time for comparison, and calculates the cross-correlation
coefficient for each comparison. (See Figure~\ref{Figure:
FT_Explainer} for an illustration.)  We assume that the wind vector is
then represented by the offset distance between the basis portion and
the comparison portion with the highest cross-correlation score. We do
not use an arbitrarily set minimum correlation score as a cutoff for
eliminating candidate wind vectors and instead choose to eliminate
spurious results after the tracking. This process is repeated
systematically throughout the mosaic. One example of using automated
tracking on \emph{Galileo} data taken during its flyby of Venus can be
found in \citet{Toigo94}.

The only user controlled parameters for this analysis are the size of
the comparison box that the search algorithm uses, and the density of
wind vectors for the final result that the user desires. These two
parameters are distinct. It is best to use the smallest box size
possible in order to minimize the position error associated with each
wind vector (see the error analysis section, Section~\ref{Section:
Error Analysis} for more discussion). However, if the basis image and
the region surrounding it are devoid of high-contrast cloud features,
the feature tracking algorithm can return inaccurate results because
there may be multiple locations where the algorithm believes there is
high correlation in a region featuring little contrast.

Our method of counteracting this problem is to use a relatively large
comparison box initially, which maximizes the presence of
distinguishing cloud features in the comparison images. The caveat
here is that using a large window will result in the algorithm
essentially measuring the offset between the distinguishing
feature(s), which can be anywhere in the comparison image. However,
the algorithm designates the location of the velocity vector to be at
the exact center of the comparison image. This can lead to
\emph{position error} in the wind velocity vectors, which can cause
errors in the calculation of vorticities or spatial derivatives of the
flow even if the magnitude and direction of the wind vectors
themselves are accurate. Therefore, in order to maximize the
localization and precision of each result, we execute the correlation
algorithm again, and using the initial large comparison box result as
a guide (i.e. an initial guess), perform the final velocity
calculation using a small comparison box. We restrict the search area
for seeking the small comparison box with the highest
cross-correlation score to a relatively small area (20x20 pixels, or
approximately 57,600 km$^2$) around the initial guess, effectively
forcing the final velocity vector to be within some range around the
initial guess. We have found this method to be especially useful in
spatially resolving areas of high horizontal shear and reducing the
chance that an artificially inflated value for vorticity (which
depends on the value of velocity shear) will result in our later
analysis. Similar techniques (of using both large and small
correlation boxes) were employed by \citet{Rossow90} in
their analysis of \emph{Pioneer} Venus images. A discussion of the
position error of our velocity vectors can be found in section
\ref{Section: Error Analysis}.

One issue to mention here is the question of what box size should be
used for the comparison images, both large and small. For the large
box, we want to keep it as small as possible to reduce position error
but maximize the chance of unique distinguishing features for each
comparison throughout the mosaic. We performed a statistical analysis
of the correlation scores returned for each velocity vector returned
by the algorithm using the box size as the only adjustable
parameter. The mean of the correlation scores increased linearly with
increasing box size at first as expected. However, the mean started to
approach a limit with larger box sizes as they reached a regime of
diminishing returns. We selected a box size of 100x100 pixels as our
large box, which is where the median correlation scores began to
asymptotically approach their peak. For the small box, we use its size
to directly control the density and localization of the results. This
can be judged subjectively by using the smallest box size that returns
as few spurious results as possible when combined with targeting from
the large box results. We performed this test using small boxes of 10,
20, and 30 pixels, and each returned similar results. Thus, we
selected the 10 pixel box to maximize resolution and minimize
positional error.

\section{Results}

Since there are three mosaics, three different combinations of image
comparisons were input into our feature tracking algorithm. The
results shown here represent an average of the three individual
comparisons. Spurious results were identified by mapping the zonal and
meridional velocities from each individual comparison as a grayscale
map and identifying pixels with obvious differences against the
surrounding region. Although we tested several methods similar to bad
pixel removal algorithms used in astronomy (in order to remove cosmic
ray hits, for example) of objectively removing spurious results
returned by the cloud trackers, none were consistent in removing all
of the inaccurate results. Therefore, we were forced to remove the
spurious results manually. It is possible that certain cloud
morphologies may have yielded spurious results that appeared to be
valid. However, we believe this is unlikely because the spurious
results were typically found in areas of bad imaging data (near image
edges, moon shadows, or other image artifacts) and that our technique
of using both large and small correlation boxes reduced the
possibility of this occurrence. In locations where results have been
removed, we use an average of two comparisons or the only valid
result. Locations with no valid data have been filled in using a
nearest-neighbor averaging algorithm. Less than 1 percent of the main
body of the maps had no valid results and were filled using
nearest-neighbor averaging; these areas were concentrated at or near
the edges of the maps.

\subsection{Wind Velocities}

The feature tracking algorithm returns separate values for offset in
the zonal and meridional directions. Velocities were determined using
the following equations:

\begin{equation} u(\lambda, \phi) = \left(x_{\mathrm{off}}(\lambda, \phi) \times d \times \left(\frac{\pi}{180}\right) \times r(\phi)\right) /\: t
\end{equation}

\begin{equation} v(\lambda, \phi) = \left(y_{\mathrm{off}}(\lambda, \phi) \times d \times \left(\frac{\pi}{180}\right) \times R(\phi)\right) /\: t
\end{equation}


where $u$ and $v$ are zonal and meridional velocity as a function of
longitude $\lambda$ and planetographic latitude $\phi$,
respectively. $x_{\mathrm{off}}$ and $y_{\mathrm{off}}$ represent the
displacements in the zonal and meridional direction in pixels, and $d$
is degrees per pixel for the mosaics ($\approx 0.01$). The offsets are
then converted into radians and then multiplied by the radii of
curvature in the appropriate direction at that point. Radii of
curvature are determined by the following equations:


\begin{equation} 
r(\phi) = R_e (1 + \epsilon^{-2}\:  \tan^2 \phi)^{-\frac{1}{2}}
\end{equation}

\begin{equation} 
R(\phi) = R_e\epsilon^{-2}\left(\frac{r(\phi)}{R_e\: \cos \phi}\right)^3
\end{equation}

where $r$ and $R$ are the zonal and meridional radii of curvature,
respectively, as a function of planetographic latitude $\phi$. The
original map coordinates for the mosaics are in planetocentric
latitude, and conversions are made appropriately. $R_e$ is the
equatorial radius of Jupiter, and epsilon ($\epsilon$) is defined as
the ratio of a planet's equatorial to polar radius ($R_e / R_p
\approx$ 1.069 for Jupiter). Finally, dividing the displacement by the
time $t$ between observations (3216 s, or twice this for the
comparison between the first and last observations) yields a result
for velocities in m s$^{-1}$. (Each observation sequence consisted of
4-6 raw images taken over a span of approximately 5 minutes, and we
assume that there was no significant motion of clouds within each
mosaic. In addition, the sequence was designed to keep a consistent
time interval between observations for raw images covering the same
area of the target.)

Figure~\ref{Figure: Wind Vectors} shows a summary of our results as a
map of wind velocity vectors. In this figure, only a ninth (every
third vector in each dimension) of the total number of vectors in our
dataset are shown for the purposes of clarity. Our feature tracking
technique is clearly able to resolve the GRS high-velocity collar, the
relatively calm central region, and jets to the south and northwest of
the GRS. We measure a maximum tangential velocity near 170 m s$^{-1}$
along the southern edge of the GRS. Our results show agreement with
previous studies: the maximum value measured using our automated
techniques is consistent (in both magnitude and location of the
velocity vector) with manual measurements on the G28 dataset made by
\citet{Simon-Miller02}, her figure 6. However, Simon-Miller also
reports velocities near or greater than 190 m s$^{-1}$ on all sides of
the high-velocity collar from her analysis. We cannot explain this
inconsistency, as large navigational uncertainties would be required
in our images to account for the error. We tested the possibility that
the discrepancy is a result of our automated tracker failing to track
very small features that are easier to track by hand and were instead
returning an averaged velocity representative of the region around the
small feature. However, this theory failed as tests with smaller
tracking box sizes did not return systematically increased velocities
in the GRS collar. Furthermore, manual cloud tracking has been
employed on the G28 dataset by \citet{Legarreta05}; they report a
maximum velocity in the flow collar of 180 m s$^{-1}$. Overall, all of
the G28 maxima lie slightly higher than the maximum manual measurement
($\sim$150 m s$^{-1}$) made by \citet{Vasavada98} on GRS imaging data
from 1996, suggesting a modest increase in the strength of the
velocity collar over a four-year period. However, we have analyzed the
G1 imaging dataset used by \citet{Vasavada98} and have performed a
similar automated feature tracking analysis on their dataset between
two of their images that are separated by nearly 1 hour. The maximum
tangential velocities as measured in the G1 dataset by Vasavada
manually and by this work automatically are in agreement, suggesting
that our automated analysis is returning the velocities properly and
that a real strengthening has occurred in that region of the collar.

Zonal and meridional velocity profiles of the GRS from our analysis
are shown in Figures~\ref{Figure: Zonal Velocity Profile}
and~\ref{Figure: Meridional Velocity Profile}. We generate the zonal
velocity profile by taking measurements within 1.5$^{\circ}$ longitude
of the central meridian of the GRS in the G28 dataset (-5.8$^{\circ}$
W) and average over 0.25$^{\circ}$ bins in latitude. The meridional
velocity profile is generated in a similar manner by taking
measurements within 1.5$^{\circ}$ latitude of 20$^{\circ}$ South
planetocentric latitude and averaging over 0.25$^{\circ}$ bins in
longitude. The targeting error of the G28 observations is shown by the
cutoff of the zonal velocity profile in the northern latitudes
compared to profiles made by previous missions. The ``double peak'' at
the southern end of the profile marks the strong zonal jet stream
located in that region. Overall, the zonal velocity profile is similar
to the profile calculated by \citet{Vasavada98}, but our analysis
shows a weakening of the northern collar countered with a
strengthening of the southern collar. These differences are relatively
substantial, with a 10-20 m/s difference in velocity on average (40-50
m/s difference at some latitudes) systematically across the collars
that is difficult to explain with the expected uncertainty ($\sim\pm5$
m s$^{-1}$, see section \ref{Section: Error Analysis}). However, we
note that \citet{Simon-Miller02} report a strengthening of both the
northern and southern GRS collar, which is inconsistent with our result.

The slight counter-rotation of the GRS core can also be seen in both
velocity profiles, confirming the result derived by
\citet{Vasavada98}. There is also increased wind shear in the G28
velocity profile, particularly in the southern half of the GRS between
the core and the high-velocity collar. Furthermore, the G28 profile
seems to be slightly asymmetrical, with the peak zonal velocity on the
southern collar roughly 30 m s$^{-1}$ higher than the highest zonal
velocity on the northern collar. (Because the northernmost portion of
the GRS was not captured because of a targeting error [see Figure
\ref{Figure: Mosaic}], there is a possibility that somewhat higher
velocities may exist farther north, but this seems unlikely given the
fact that the zonal velocity profile in Figure \ref{Figure: Zonal
Velocity Profile} appears to begin turning over.) The asymmetry is not
as pronounced for the meridional velocities, with a difference of
about 10 m s$^{-1}$ between the peak velocities on the western and
eastern edge of the velocity collar.

\subsection{Turbulent Eddies}

The remarkable high resolution of our images enables us to map the
central region of the GRS with great detail. A full-resolution wind
velocity vector map of the region is shown in Figure~\ref{Figure:
Central Wind Vectors}. Most of the central region appears to be
incoherent and turbulent, but a couple of notable features can be
seen. An anti-cyclonic sub-vortex that is likely fed by the
high-velocity collar can be seen to the northeast, and hints of
another sub-vortex can also be seen in the southwest
corner. Furthermore, the northern half of a cyclonic circuit (flowing
W to E) can be seen near the geographic center of the GRS. There also
appears to be a branch from the main high-velocity collar that feeds
this motion. The southern half of the circuit is not clearly seen in
this figure and appears to be lost in the chaotic
motion. \citet{Sada96} found a small, cyclonic vortex near the center
of the GRS from their analysis of \emph{Voyager} images, whereas
\citet{Vasavada98} found a complete cyclonic circuit surrounding the
center of the GRS. The fickle nature of the velocity features in the
GRS central region based on observations over a span of over 20 years
only reaffirms the chaos and turbulence that likely dominate this
region. The timescale over which these features change is unknown, but
it is probably less than a couple of years.

\subsection{Particle Trajectories}

We treated the problem of integrating velocities to create particle
trajectory paths as an ODE and composed an algorithm that used the
velocity fields and the classical fourth-order Runge-Kutta (RK) method
to solve for the trajectories. Our algorithm was written using steps
outlined in \emph{Numerical Recipes in C} \citep{NR}, but using IDL
instead. The algorithm starts its integration with a user-controlled
input point of origin. The fourth-order RK algorithm requires
derivatives of the velocity fields in both horizontal directions for
its calculations (i.e. a calculation of the horizontal wind
shear). Because the wind shear calculated using the raw, noisy
velocity data can lead to artificially inflated numbers for the shear
and negatively affect later calculations, we smoothed the velocity
field in a 1 x 1$^{\circ}$ box centered at each trajectory point, and
then calculated the shear using the smoothed field. We set our
smoothing functions to be velocity functions that vary only linearly
in latitude and longitude. Once a new trajectory point was calculated
using the integrator, the algorithm repeats the smoothing using a box
centered at the new point.

Figure~\ref{Figure: Trajectories} shows our map of the representative
trajectories we have selected for use in our analysis. None of the
calculated trajectories yielded closed curves; in fact, all of them
spiraled into the center of the GRS. We sought to investigate why the
trajectories were not fully closing on themselves. We first tested our
integrator on a synthetic dataset that would return a smooth, closed
circle as its trajectory. Once we verified that the integrator worked
properly, we added noise to our dataset that would affect the final
trajectory in order to assess whether errors in the dataset or in the
analysis were affecting the trajectories. We discuss the various
sources of errors and how the gaps in the trajectories affect our
interpretation of the results later in Section \ref{Section: Error
Analysis}. However, through extensive testing of the synthetic dataset
with noise, we are confident in concluding that the various sources of
error in the dataset and in our analysis can only explain a small
fraction of the gap in the particle trajectories. We speculate that
transient eddies, i.e.~a time-dependent flow that is present in the
GRS high-velocity collar such as Rossby waves or other pulsations,
were captured in our observations and are the reason behind the
trajectories' failure to close. Although it is a possibility that the
convergence of the trajectories is a real result, it seems unlikely
given that most dynamical models of the GRS predict a divergent flow
over the GRS that would occur over a much longer timescale than what
is seen from our trajectories \citep{Conrath81}.

\subsection{Vorticities}

Relative vorticities were determined using the following equation
\citep{Dowling88}:

\begin{equation} 
\zeta = -\frac{1}{R}\:\frac{\partial u}{\partial \phi} +
\frac{u}{r}\:\mathrm{sin}\:\phi + \frac{1}{r}\:\frac{\partial
  v}{\partial \lambda} 
\end{equation}

Absolute vorticities were determined by simply adding the Coriolis
parameter $f = 2 \Omega \sin \phi$ to the relative vorticities.  The
vorticity calculation also requires the derivatives of the velocity
field. We smooth the velocities over a 1 x 1$^{\circ}$ box centered at
each velocity vector map gridpoint using the same smoothing algorithm
mentioned earlier to calculate trajectories, and then use those wind
shear values to calculate vorticities.

Figure~\ref{Figure: Relative Vorticity} shows a grayscale map of
relative vorticity for the Great Red Spot and surrounding
region. Lighter shades indicate positive (anti-cyclonic or
counter-clockwise in the southern hemisphere) vorticity whereas darker
shades signify negative (cyclonic) values. It can be seen that the
high-velocity collar is strongly anti-cyclonic. Curiously, the
southern half of the high-velocity collar appears to have larger
magnitude vorticity values compared with the northern half. However,
the most interesting feature seen in Figure~\ref{Figure: Relative
Vorticity} is the ring of cyclonic vorticity that surrounds the GRS
high-velocity collar. Similar rings surrounding the GRS have been
observed before from 5-$\mu$m infrared spectroscopy \citep{Flasar81},
but this observation marks what we believe to be the first such
observation of a ring from a dynamical perspective. Although
\citet{Dowling88} calculated a relative vorticity map for their
analysis of \emph{Voyager} images, no clear ring of cyclonic vorticity
can be seen in their figure. Our result for the relative vorticity
based on the G1 dataset is also shown in Figure~\ref{Figure: Relative
Vorticity}. Because the cyclonic ring is also seen in the G1 dataset,
we can conclude that this feature is not transient but appears to be a
distinguishing characteristic of the Great Red Spot.

Although the trajectories that we have calculated are open, we present
an analysis of how absolute vorticity changes along these
trajectories. The original trajectories were divided into a locus of
points at 30-minute intervals along each trajectory. We then
calculated the vorticity at each point using a modified version of our
algorithm by performing a separate velocity smoothing around each
trajectory point. (Thus, we do not perform an interpolation of the
vorticity map at the selected trajectory points. Instead, we perform a
separate calculation and smoothing at each point.) Three trajectories
were selected for analysis, and the result is shown in Figure
\ref{Figure: Vorticity vs. Latitude}.

Potential vorticity (PV) is defined as $\frac{(\zeta + f)}{H}$ where
$(\zeta + f)$ is absolute vorticity and $H$ is a measure of the layer
(or pressure) thickness of the fluid layer. This is a form of
potential vorticity first developed by \citet{Ertel42} and also used
in \citet{Dowling88}. Therefore, relative changes in the layer
thickness of the rotating fluid layer can be tracked by analyzing how
absolute vorticity changes along a trajectory, assuming conservation
of PV. Figure \ref{Figure: Vorticity vs. Latitude} generally matches
the results determined by \citet{Dowling88}: layer thickness
variations as a function of latitude cannot be solely attributed to
the $\beta$ effect (the effect of the Coriolis parameter varying with
latitude). In particular, trajectory F shows a thickening of the fluid
layer over a narrow range of latitudes from the distinct slope in
Figure \ref{Figure: Vorticity vs. Latitude} compared with the slope of
the Coriolis parameter $f$. Trajectories A and C show similarities
overall in their slopes in comparison to $f$, but portions of the
trajectories also show distinct slopes. Similar results are shown in
\citet{Dowling88}, their Figure 14. One caveat to our analysis is that
because the calculated trajectories spiral in towards the GRS center,
our assumptions for time-independent flow and PV conservation are not
entirely valid, and the effect of this spiraling will be discussed in
section \ref{Section: Error Analysis}.

\subsection{Shape and Aspect Ratio}

Analysis by \citet{Simon-Miller02} using photometric analysis of
records dating back to 1880 and images from spacecraft flybys has
shown a decrease in length of the GRS by nearly 50 percent. We wish to
supplement what is known about this particular aspect of the GRS by
analyzing the particle trajectories that were calculated using our
data and by \citet{Dowling88}. We measured the length and width of
trajectory curves from both datasets by defining two axes that spanned
each trajectory curve: one placed along the latitude halfway
in-between the minimum and maximum latitudes of each trajectory, and
another placed along a line of longitude in a similar fashion. The
distance of the line segment between the intersections of the
trajectory points and the axes was then measured. We measured the
length and width of the trajectory shapes from \citet{Dowling88},
their Figure 3, using an electronic reproduction of their figure and
Adobe Photoshop. This reproduction and our annotations marking which
trajectory shapes we measured are shown in \ref{Figure: Dowling
Fig3}. From our dataset, trajectory A was not analyzed because the
upper portion of the trajectory was not calculated due to that portion
of the GRS not being imaged.

Our results are summarized in Table \ref{Table: Aspect Ratios}. We
report there is a slight increase in the aspect ratio (which we define
as length divided by width) for the outermost GRS trajectories
(trajectories 1-3 and B-C) between the \emph{Voyager} and
\emph{Galileo} missions. This is somewhat surprising as a 3-5$^{\circ}$
decrease in the length of the GRS as estimated by
\citet{Simon-Miller02} over the twenty years between the missions
would correspond to a decrease in the aspect ratio by around 20 to 25
percent (assuming the same width). In contrast, the aspect ratio of
the trajectories seems to have decreased for the more interior
trajectories (trajectories 5-8 and D), which is in step with the slow
circularization of the GRS with time. However, we note that since our
trajectories spiral inward, the true aspect ratios for our trajectory
curves are likely slightly smaller, and that our values represent an
upper boundary. We also find that the aspect ratio of the trajectories
from the \emph{Voyager} data increases as the radius of the
trajectories decreases; curiously, the opposite is true in the
\emph{Galileo} data. It is unclear if this is a manifestation of a
physical change of the flow pattern and shape of the GRS or a change
caused by a difference in the methods used to interpolate and smooth
velocities and integrating them to get trajectories. We would also
like to note a different line of evidence showing the ``shrinking'' of
the GRS with time: Figure \ref{Figure: Meridional Velocity Profile}
distinctly shows a clear decrease in the length of the GRS, by
measuring the distance between the peak meridional velocities on the
western and eastern sides of the vortex.

\section{Error Analysis}
\label{Section: Error Analysis}

There are three main categories of error that affected our
results. The first category is uncertainties in the location of the
cloud features in the image mosaics. These uncertainties are
introduced during image processing and mosaic creation when tiepoints
(common features) used to stitch together the raw images are
identified and selected. The second category is errors in the location
of the velocity vectors. This error is generated by our automated
cloud feature tracking algorithm. Although the velocity vectors are
produced in a regular grid, the particular cloud feature that was
tracked and was responsible for generating the velocity result does
not necessarily correspond with the \emph{exact} location of the
velocity vector but can be slightly offset. The final category is
errors made by the automated feature tracker from the inherent
variability in Jupiter's cloud morphology. If the a particular feature
changes sufficiently between observations, then there is a good chance
of the algorithm failing to recognize the correct feature by selecting
the incorrect correlation peak and assigning an inaccurate velocity
vector. Unfortunately, this type of error is difficult to quantify but
can usually be recognized by spurious results returned by the
algorithm. However, the short time interval between each G28
observation helped to reduce this error greatly at the expense of
reduced resolution in velocities.

When tiepoints for the images were selected during mosaic
construction, we ensured that the maximum uncertainty in cloud
position would be less than 1 pixel by repeating the tiepoint
selection process until this criterion was met. (The NAV2 subroutine
under VICAR calculates tiepoint errors by performing an area
correlation of the image around each tiepoint.) Thus, by reducing the
uncertainty as much as possible during the mosaic creation process, we
ensured that the errors in velocities calculated using our feature
tracking algorithm would be kept to a minimum. We assume for the
purposes of this error analysis that the overall uncertainty in cloud
position is estimated to be 0.5 pixel for an entire mosaic. We can
then expect an error in cloud offsets from the feature tracker to be
$\sim$1 pixel on average. The error in cloud offsets is not expected
to be greater than 2 pixels. Propagating a 1 pixel uncertainty in raw
offset values yields an uncertainty of about $\pm$3.5 m s$^{-1}$ in
velocity. The maximum uncertainty in our velocity measurements is
estimated to be about $\pm$7 m s$^{-1}$. Errors are dependent on the
radii of curvature but do not vary by more than $\pm$0.2 m s$^{-1}$.

Another source of error is error in the \emph{location} of the
calculated velocity vectors. This error is directly controlled by the
size of the feature tracking correlation box, and is also dependent on
the amount of horizontal shear. While the velocity vector's origin is
assigned to be located at the center of the correlation box, the
particular cloud feature that the feature tracking algorithm locks
onto and uses for calculating the cloud motion can be located anywhere
within the correlation box. Our choice of a 10 pixel (120 x 120 km)
box is sufficient to keep the error to near $\pm$2 m s$^{-1}$ using typical
values for horizontal velocity shear in Jupiter's atmosphere. If we
had used a 30 pixel (360 x 360 km) box, we estimate the errors to be
nearly 6 - 7 m s$^{-1}$. Thus, we estimate the overall error in the
velocities to be typically around 4 m s$^{-1}$.

This error is insufficient to explain the gaps that are seen in the
calculated trajectories, because we determined that the combined
errors from camera pointing and the positions of the velocity vectors
would be insufficient if they were distributed randomly throughout the
synthetic dataset. Furthermore, we tested how the integrator would
behave if the errors were correlated over a length scale similar to
length scales found in the imaging dataset. We reason that it would be
possible for errors to be similar in magnitude over a length scale of
typical cloud features. This ``correlated error'' would produce a
deviation in the trajectory that would constructively add, whereas
randomly distributed errors would tend to contribute random-walk like
behavior and would presumably produce little net effect on the
trajectories. We tested our theory by performing an FFT on one of the
image mosaics to create a power spectrum as a function of length. We
determined that the dominant length scale for the FFT on our image
mosaics was on the order of a few degrees. The phase of the FFT was
then randomly scrambled and then inverted to produce a noisy image
with the same length scale as the original data. This noisy image was
then scaled to be in proportion with the expected error we would
obtain from both camera pointing and velocity vector position error,
and added to the synthetic dataset. Testing with the integrator on
this experimental data resulted in minor trajectory gaps, but was not
sufficient to explain the wide gaps in our original trajectories. We
therefore conclude that some natural variation in the GRS velocity
collar is responsible for the trajectory gaps that result in our
data. We have no evidence that clouds at different heights adversely
affected our results, but this could be another source for error.

Because the calculated trajectories do not fully close in on
themselves, we cannot make the assumption that they are fully
time-independent. Thus, conservation of potential vorticity may not be
fully applicable to these trajectories because they are not true
trajectories that track a single air parcel. Instead, the paths that
we have calculated most likely traverse between multiple parcels. We
estimate the typical gap between the starting and ending points of our
trajectories to be $\sim$1$^{\circ}$. Using the error estimation
methods outlined in \citet{Dowling88}, we estimate the errors in the
derived thicknesses to be at the 30\% level. (For comparison,
\citet{Dowling88} calculated their error to be at around the 3\%
level.) Also, we would like to note that we executed the path
integrator using the velocity data that we calculated using the G1
imaging dataset with similar results: all of the trajectories around
the GRS failed to close by a similar margin. Although this is a
non-negligible error, we remain confident that our analysis shows that
the GRS has not undergone any drastic changes in its dynamics and
overall structure since the \emph{Voyager} era, and that our overall
conclusions remain valid.

\section{Discussion and Conclusions}
\label{Section: Discussion}

It is not expected that this paper will be able to resolve the
question of whether it is inherently better to use an automated cloud
feature tracking system over a manual or user-controlled system, but
we believe we have conclusively demonstrated that an all-automated
approach can return quality results. The automated approach reduces
the time required to painstakingly track clouds by hand and eye,
though there is some additional time required to eliminate any
spurious results. The automated method is particularly useful for
returning a high-density, regular grid of velocity vectors, something
that is not achievable with a manual approach that will typically
result in a scattered grid of velocity vectors. Furthermore, the
high-density velocity vector grid facilitates the calculation of
particle trajectories and vorticities. However, the usefulness of an
automated approach is most likely constrained by the native image
resolution of the imaging dataset. It would be interesting to measure
what minimum image resolution is necessary for an automated approach
to be useful. It should also be noted that there is a cutoff in the
time gap between observations where manual cloud feature tracking will
become more useful. Using the feature tracking algorithm on G1
observations where the time gap between observations was 9-10 hours
yielded a high percentage of spurious results and were unfit for
analysis. Presumably, the cloud morphology had sufficiently evolved
between those observations so that the tracking algorithm failed to
recognize clouds at the later observation. However, the presence of
compression artifacts in the G1 data may have negatively influenced
the results as well. Overall, we conclude from our datasets that an
automated analysis is ideal for high-resolution observations with
relatively short ($\sim$1-2 hours) timesteps between images, whereas
manual tracking is better for observations of any resolution but with
much longer timesteps between observations. However, an exception for
when longer timesteps are better suited for automated tracking may be
for capturing subtle details such as eddy momentum fluxes
\citep{Salyk06}.

The most interesting feature that resulted from our analysis is the
ring of cyclonic vorticity that surrounds the high-velocity collar of
the Great Red Spot. A similar feature surrounding the high-velocity
collar has been observed spectroscopically in 5$\mu$m
\citep{Flasar81}, but ours is presumably the first such observation
made from a dynamical perspective. We could not conclusively find a
cyclonic ring feature in the results of \citet{Dowling88} and
\citet{Sada96}; however, this feature was detected in our re-analysis
of \emph{Galileo} G1 data first analyzed by \citet{Vasavada98}. Thus,
this feature is a real phenomenon and not an artifact of our dataset
or our analysis techniques. We believe that it is unlikely that this
feature is transient in nature, and we attribute the failure in
detection from previous studies to the limitations brought by manual
cloud tracking and the low resolution of past images. Thus, the
structure of the Great Red Spot extending outward from its center can
be summarized as a slightly cyclonic central core, a strongly
anti-cyclonic high-velocity collar, and a narrow cyclonic outer
ring. The outer cyclonic ring itself may not be interesting, as one
could expect a cyclonic component to the vorticity of the GRS as the
winds decay away from the center of the vortex. This is exemplified in
Equation 4.9 from \citet{Holton04}:

\begin{equation} \zeta = -\frac{dV}{dn} + \frac{V}{R} \label{Equation: Holton}
\end{equation}

where $\zeta$ is relative vorticity, $V$ is tangential velocity
(defined positive), $n$ is a radial coordinate (positive inward for
the GRS), and $R$ is radius of curvature (positive for the GRS). Thus,
as the tangential velocity weakens with distance away from the GRS
collar, the shear contribution of equation \ref{Equation: Holton} (the
first term) would make $\zeta$ negative. However, if the vortex winds
decayed slowly away from the center over the length scale of the GRS,
the curvature contribution of the vorticity (the second term, which is
positive in this case) would dominate, and the vorticity of the GRS
would gradually transition into the background vorticity of the
regional flow. Thus, the narrow width and strong amplitude of the
cyclonic ring signifies that the GRS winds decay rapidly with distance
away from the main flow collar. It is unclear what process is forcing
the ring to be remarkably narrow, especially when compared to the
width of the main anti-cyclonic velocity collar. Overall, we believe
the properties of the cyclonic ring around the GRS to be a useful
constraint on the dynamics of the GRS.

We also hypothesize that there could be a thermally indirect
circulation occurring in the GRS, where downwelling motion is
associated with a region of cyclonic vorticity; this type of
circulation model was first proposed by \citet{Conrath81}. If there is
downwelling motion in the cyclonic ring, it would coincide well with
the observation that this region is bright in 5 $\mu$m, as heat from
the Jovian interior could more readily escape through this outer
ring. Because the 5-$\mu$m bright ring appears to be a stable feature
around the GRS (see also \citet{Carlson96}), and because the exact
mechanism generating this feature is unknown, it is a natural target
for future studies on the stability and dynamics of the Great Red
Spot. Previous numerical modeling attempts of the GRS have been
successful \citep{Cho01} in reproducing particular features of the
wind flow, but little attention has been paid to cyclonic rings in GRS
modeling studies. \citet{Marcus04} has shown that cyclonic rings are
present around White Oval-like vortices in his numerical studies.
Furthermore, \citet{Showman06} has demonstrated that cyclonic rings
form around Jovian anticyclones as a secondary result of his study
testing the hypothesis that the zonal jets on Jupiter are powered by
thunderstorms. Thus, we believe that investigating this new ring
feature requires a numerical modeling approach in an attempt to
reconcile it with previous GRS models and in order to understand how
the properties of the Jovian atmosphere and the GRS generate this
feature and regulate its width and amplitude.

Although our analysis of vorticity change along trajectories is
imperfect because of the failure of the trajectories to close on their
point of origin, our analysis strongly suggests that the structure of
the Great Red Spot remains largely unchanged from the \emph{Voyager}
era. Our velocity measurements confirm the strengthening of the
high-velocity collar since the \emph{Voyager} era, and even since four
years prior to the G28 observations, but not to the extent previously
measured by \citet{Simon-Miller02}. Nevertheless, we hope to employ
our automated feature tracking techniques to other high-resolution
imaging datasets in the future in order to supplement our knowledge on
the solar system's diverse planetary atmospheres.

\acknowledgments
This paper benefited from the helpful comments and reviews made by
Dr.\ Anthony del Genio and an anonymous reviewer. We thank Dr.\ Aswhin
Vasavada for providing his projected \emph{Galileo} G1 imaging data
for analysis, and we would also like to thank Dr.\ Tim Dowling for
providing his results from \emph{Voyager} imaging analysis for
comparison. We thank Ossama Othman for the development and use of his
MaRC map projection software while at Cornell University. This work
was supported by NASA's Planetary Atmospheres Program and by Cornell
Presidential Research Scholars.

\bibliography{my_paper_icarus.bib}
\bibliographystyle{elsart-harv}


\clearpage	

\begin{table}
\begin{center}
\begin{tabular}{|c|c|c|} \hline

Dataset & Trajectory & Aspect Ratio \\ \hline
& 1 & 2.038 \\ \cline{2-3}
& 2 & 1.999 \\ \cline{2-3} 
& 3 & 2.020 \\ \cline{2-3}
& 4 & 2.075 \\ \cline{2-3}
& 5 & 2.143 \\ \cline{2-3}
\raisebox{1.5ex}[0pt]{\emph{Voyager}} & 6 & 2.249 \\ \cline{2-3}
& 7 & 2.336 \\ \cline{2-3}
& 8 & 2.467 \\ \cline{2-3}
& 9 & 2.678 \\ \cline{2-3}
& 10 & 3.076 \\ \hline \hline
& B & 2.179 \\ \cline{2-3}
\emph{Galileo} & C & 2.178 \\ \cline{2-3}
& D & 2.043 \\ \hline

\end{tabular}
\caption{Aspect ratios (length/width) for selected trajectories from the
  \emph{Voyager} imaging data \citep{Dowling88} and \emph{Galileo}
  data. The trajectories are illustrated in Figures \ref{Figure:
  Trajectories} and \ref{Figure: Dowling Fig3}.
  }
\label{Table: Aspect Ratios}
\label{lasttable}
\end{center}
\end{table}

\clearpage


\begin{figure}[htb]
  \centering
  \includegraphics[width=5.5in, keepaspectratio=true]{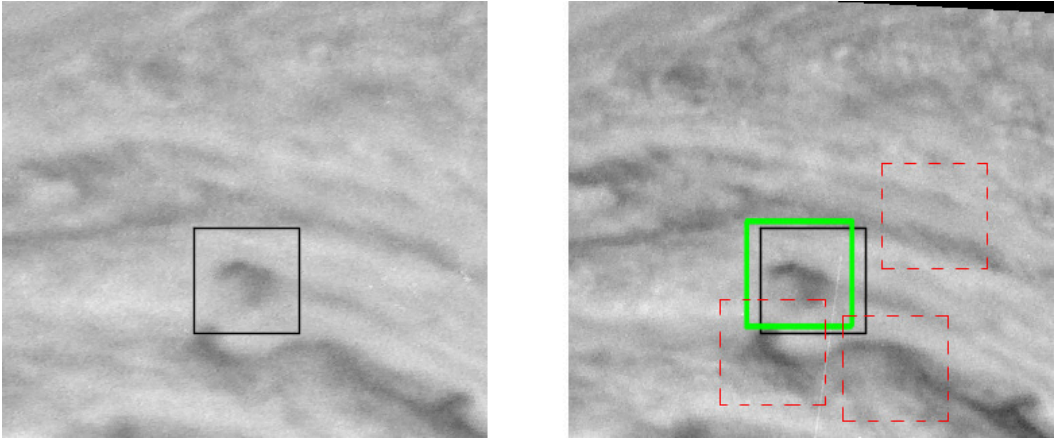}
  \caption[GRS FT Explainer]{
    \label{Figure: FT_Explainer}
    Illustration depicting how our feature tracking algorithm
    works. The code extracts a basis portion of an image at some
    initial time (left). Then, it extracts numerous comparison
    portions from the image at a later time (right), and calculates
    the one with the highest cross-correlation score (light gray,
    right) and rejects those with lower cross-correlation scores
    (dashed boxes, right). The offset distance between the light gray
    box and the black box is the basis for the calculation of wind
    velocities.  }
\end{figure}

\begin{figure}[htb]
  \centering
  \includegraphics[width=7in, angle=270, keepaspectratio=true]{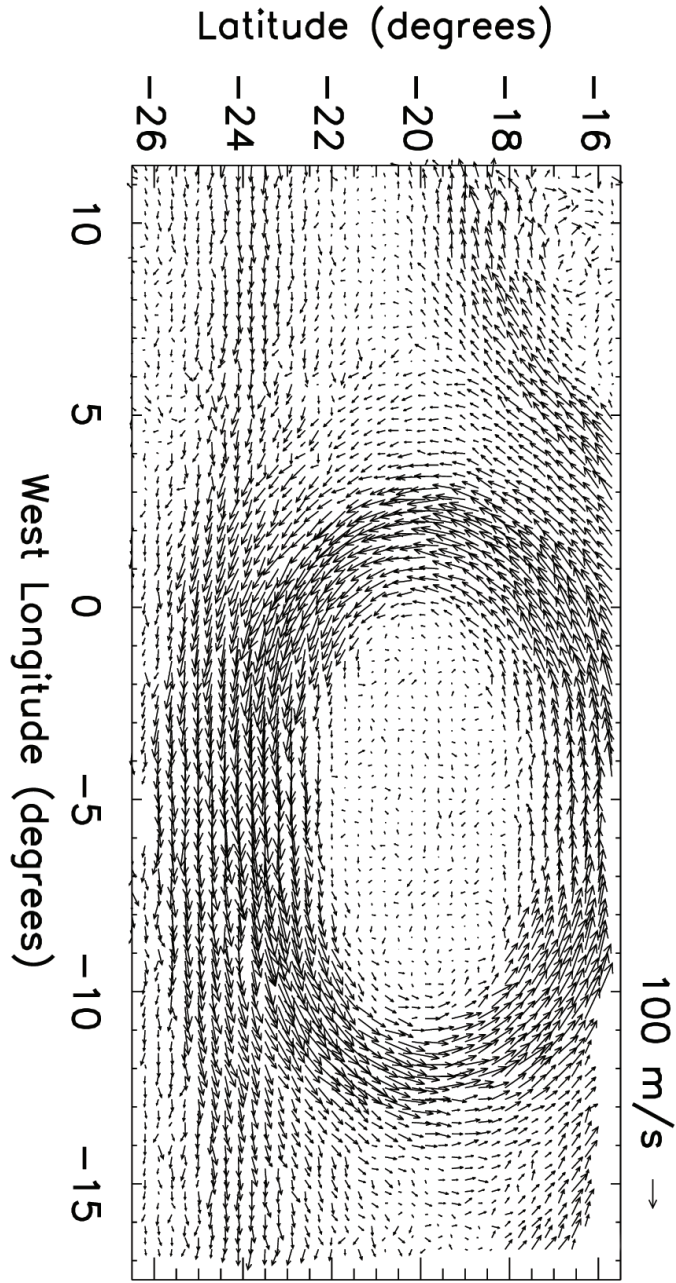}
  \caption[GRS Wind Velocity Vectors]{
    \label{Figure: Wind Vectors}
    Wind velocity vector maps of the GRS and surrounding region. Only
    a ninth of the total number of velocity vectors calculated using
    our technique are shown in this figure for the sake of
    clarity. Note the scale vector at top right.
    }
\end{figure}

\begin{figure}[htb]
  \centering
  \includegraphics[width=6in,keepaspectratio=true]{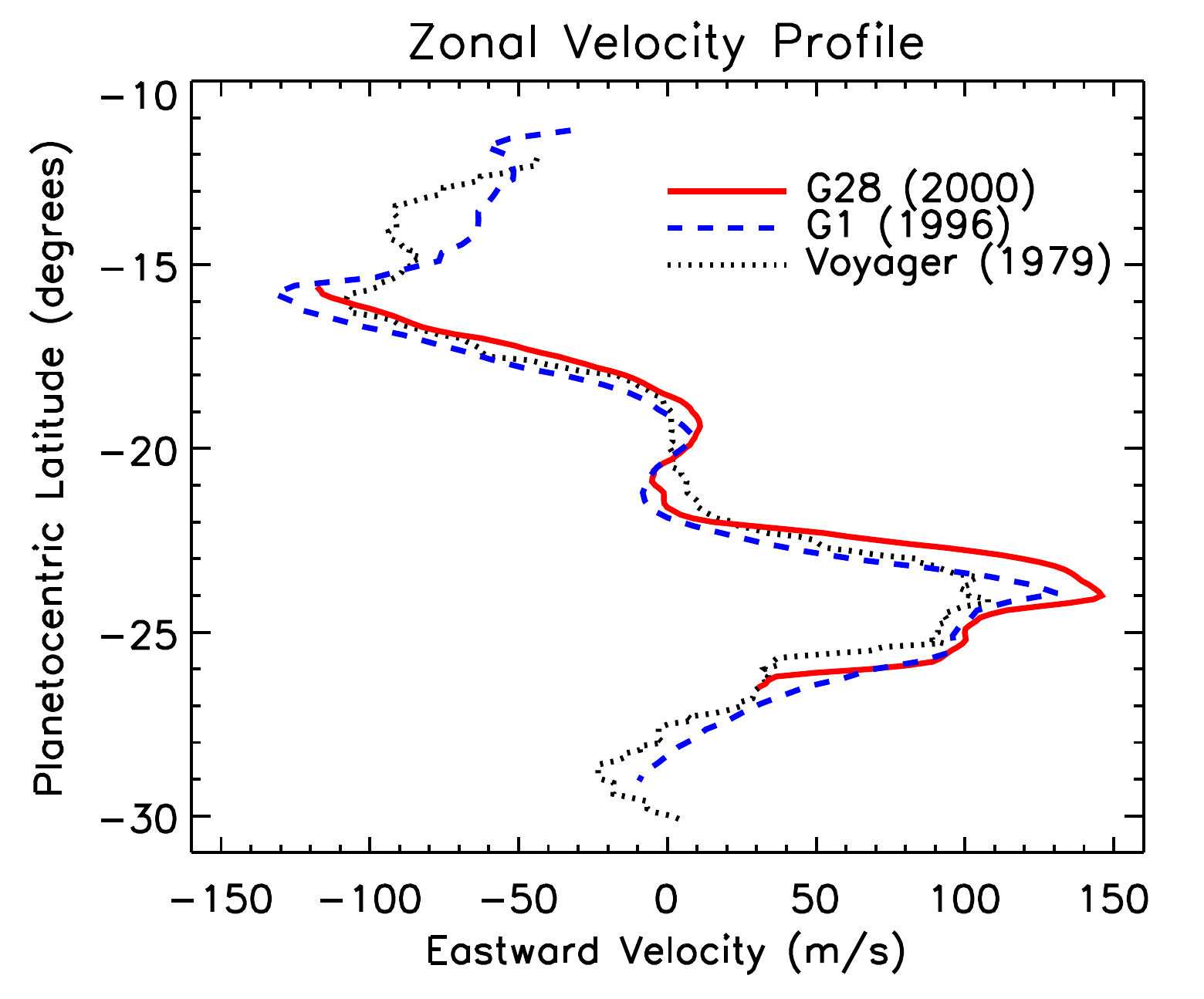} 
  \caption[Zonal Velocity Profile]{
    \label{Figure: Zonal Velocity Profile}
    Zonal velocity profile of the Great Red Spot. The solid, red line
    is calculated from our \emph{Galileo} data. Our measurements
    within 1.5$^{\circ}$ of the GRS central meridian were averaged
    over 0.25$^{\circ}$ latitude bins. The zonal velocity profile from
    \citet{Vasavada98} (whose velocity profile was calculated in the
    same manner) is shown for comparison as a dashed, blue line. The
    dotted, black line is \emph{Voyager} data from \citet{Dowling88},
    where we have converted their data into planetocentric latitude
    and averaged their data over 1$^{\circ}$ latitude bins.}
\end{figure}

\begin{figure}[htb]
  \centering
  \includegraphics[width=6in,keepaspectratio=true]{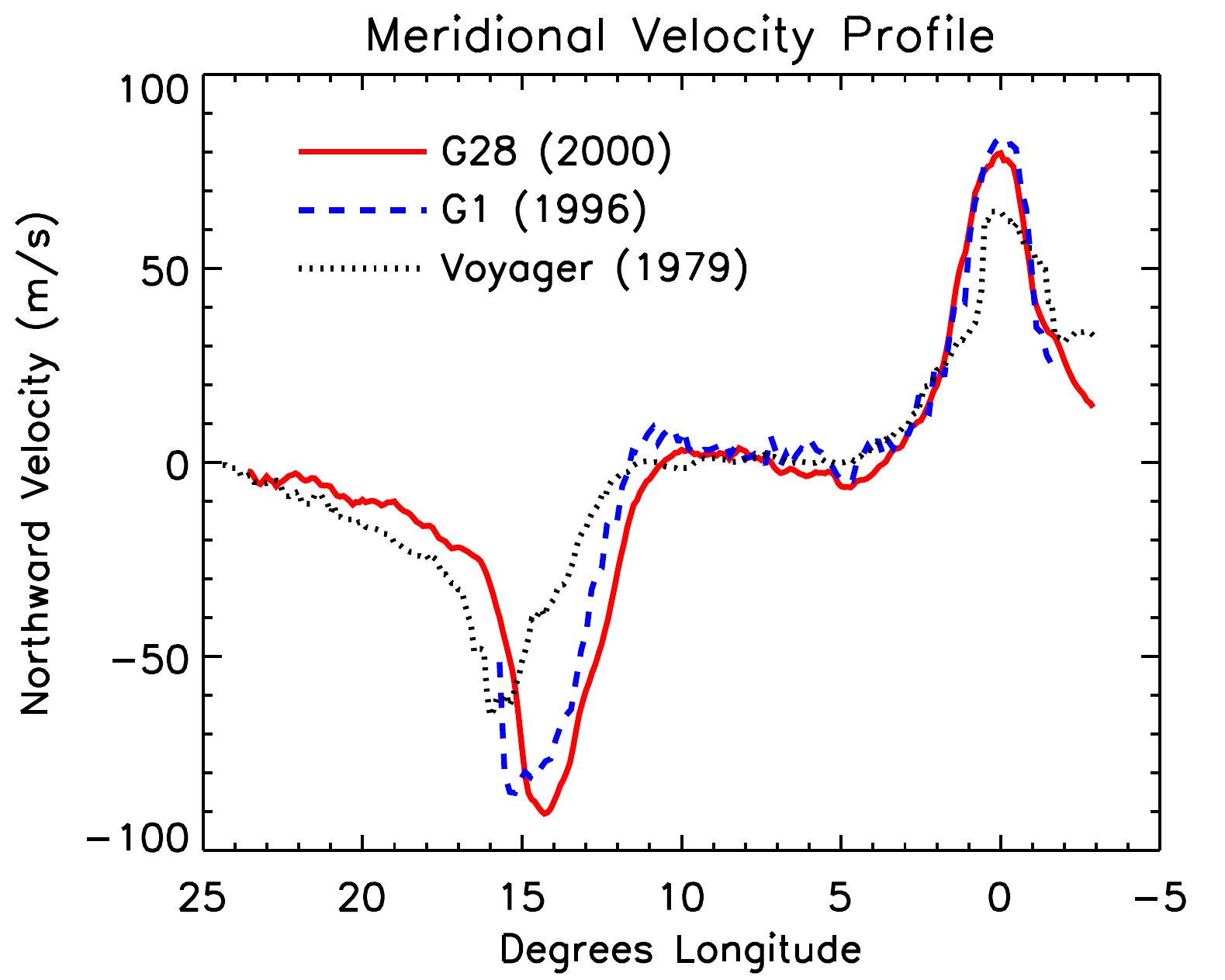} 
  \caption[Meridional Velocity Profile]{
    \label{Figure: Meridional Velocity Profile}
    Same as Figure \ref{Figure: Zonal Velocity Profile}, but showing a
    meridional velocity profile of the Great Red Spot. Measurements
    within 1.5$^{\circ}$ of 20$^{\circ}$ S were averaged over
    0.25$^{\circ}$ longitude bins for the G28 and G1 data, and over
    1$^{\circ}$ longitude bins for the \emph{Voyager} data. The axis
    for longitude is provided as a scale reference; each profile's
    zero longitude was set at the peak northernmost velocity on the
    eastern side of the GRS.}
\end{figure}

\begin{figure}[htb]
  \centering
  \includegraphics[width=7in,angle=270, keepaspectratio=true]{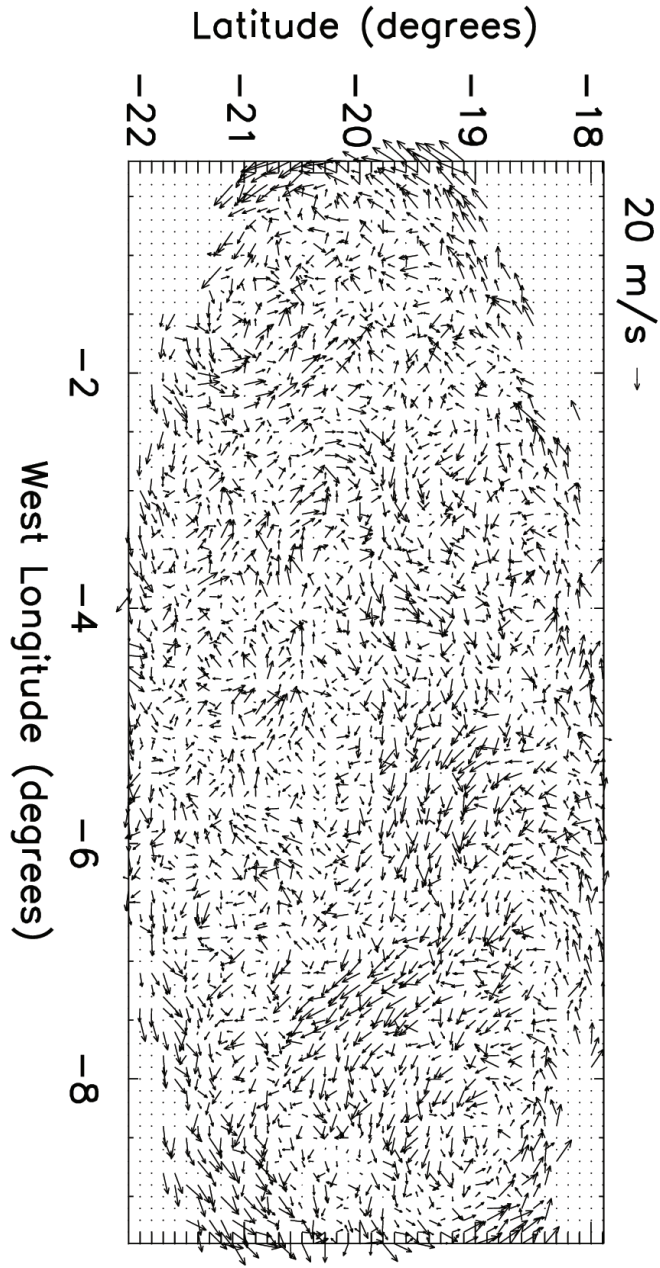}
  \caption[GRS Center Wind Velocity Vectors]{
    \label{Figure: Central Wind Vectors}
    Wind velocity vector maps of the center of the GRS. All velocity
    vectors in our dataset are shown for highest resolution. Note the
    scale vector at top left. 
  }
\end{figure}

\begin{figure}[htb]
  \centering
  \includegraphics[width=8in, angle=270, keepaspectratio=true]{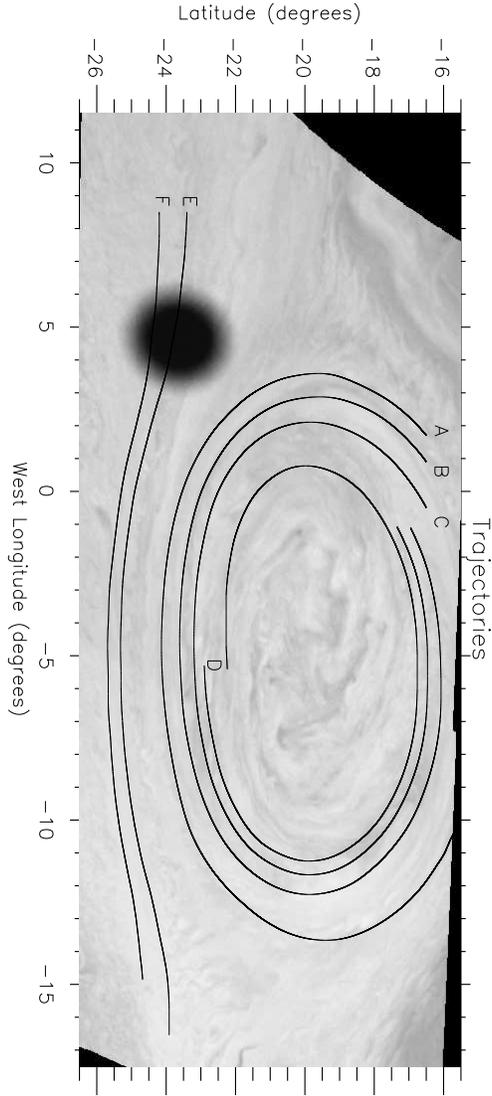}
  \caption[GRS Trajectories]{
    \label{Figure: Trajectories}
    Map of particle trajectories used for analysis, overlain on one
    of the G28 image mosaics. 
    }
\end{figure}

\begin{figure}[htb]
  \centering
  \includegraphics[height=6in, keepaspectratio=true]{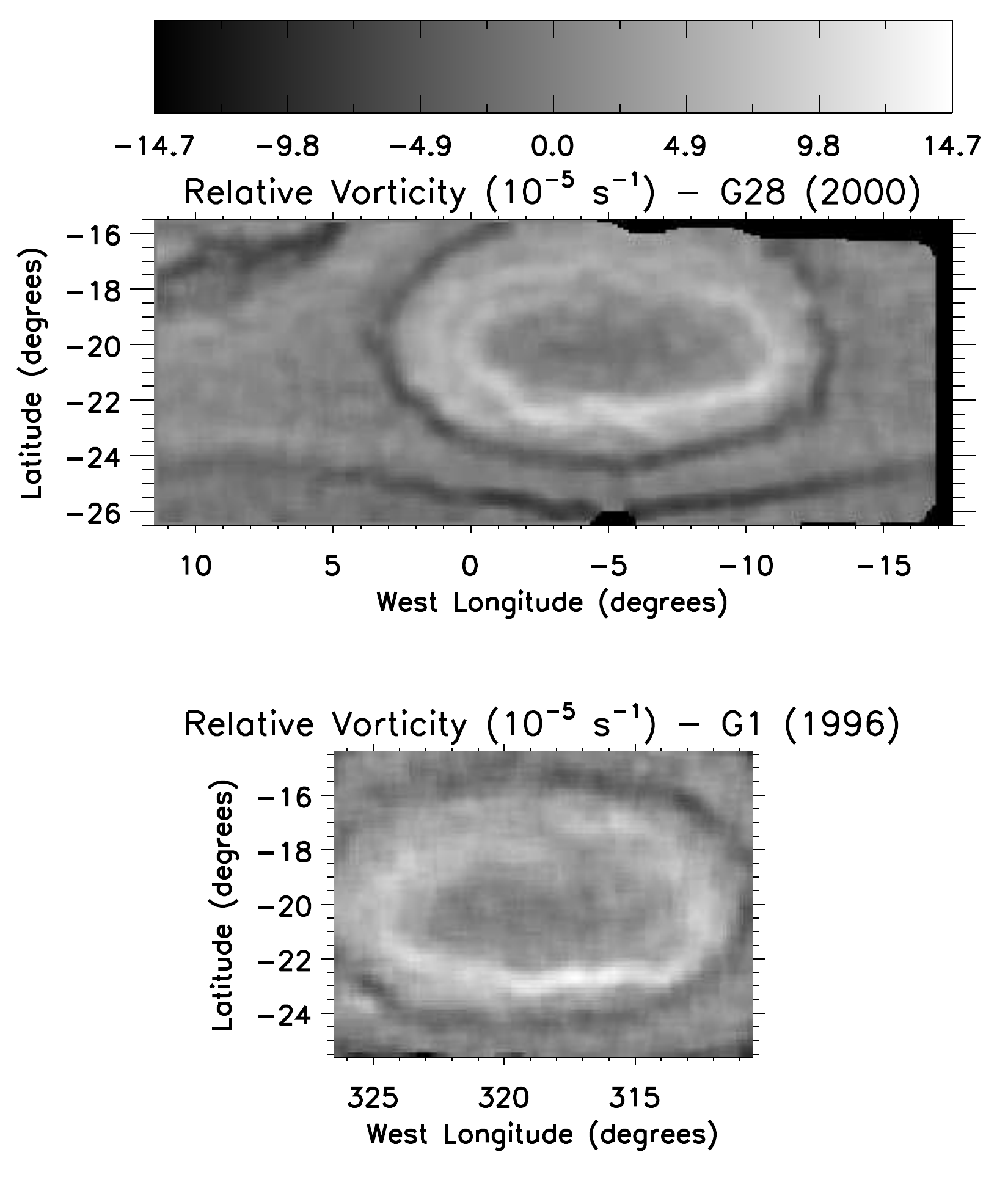}
  \caption[GRS Relative Vorticity]{
    \label{Figure: Relative Vorticity}
    Grayscale map of relative vorticity for the Great Red Spot from
    the G28 and G1 datasets. Vorticity is mapped in units of 10$^{-5}
    s^{-1}$. The Great Red Spot is in the southern hemisphere of
    Jupiter; thus positive values denote anti-cyclonic
    (counter-clockwise) motion, while negative values signify cyclonic
    (clockwise) motion.  }
\end{figure}

\begin{figure}[htb]
  \centering
  \includegraphics[width=6in,keepaspectratio=true]{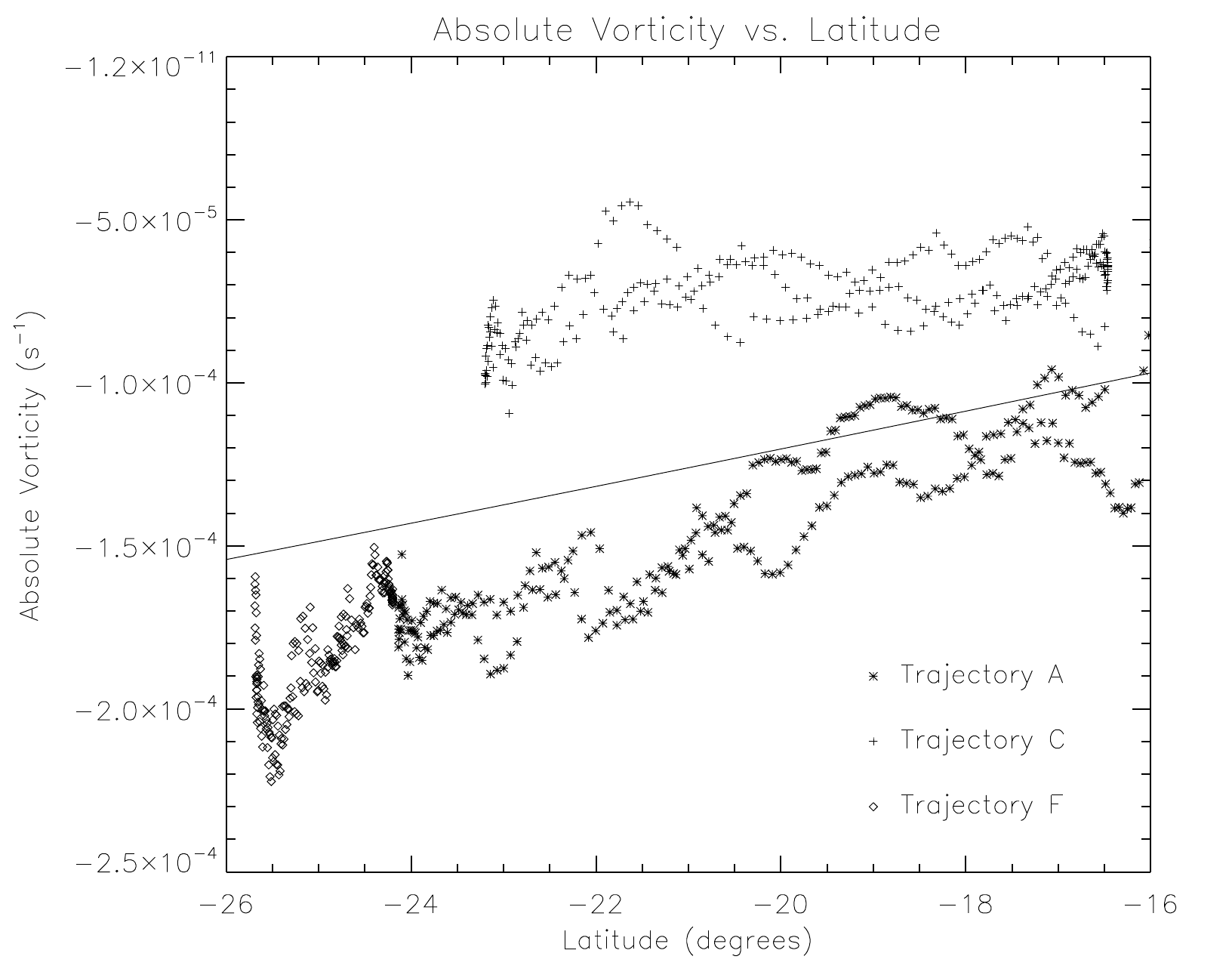}
  \caption[GRS Relative Vorticity]{
    \label{Figure: Vorticity vs. Latitude}
    Plot of absolute vorticity $(\zeta + f)$ as a function of latitude
    along three trajectories. The different plot symbols shown in the
    figure correspond to different trajectories depicted in Figure
    \ref{Figure: Trajectories}. The solid line corresponds to the
    Coriolis parameter $f$.  }
\end{figure}

\begin{figure}[htb]
  \centering
  \includegraphics[width=6in,keepaspectratio=true]{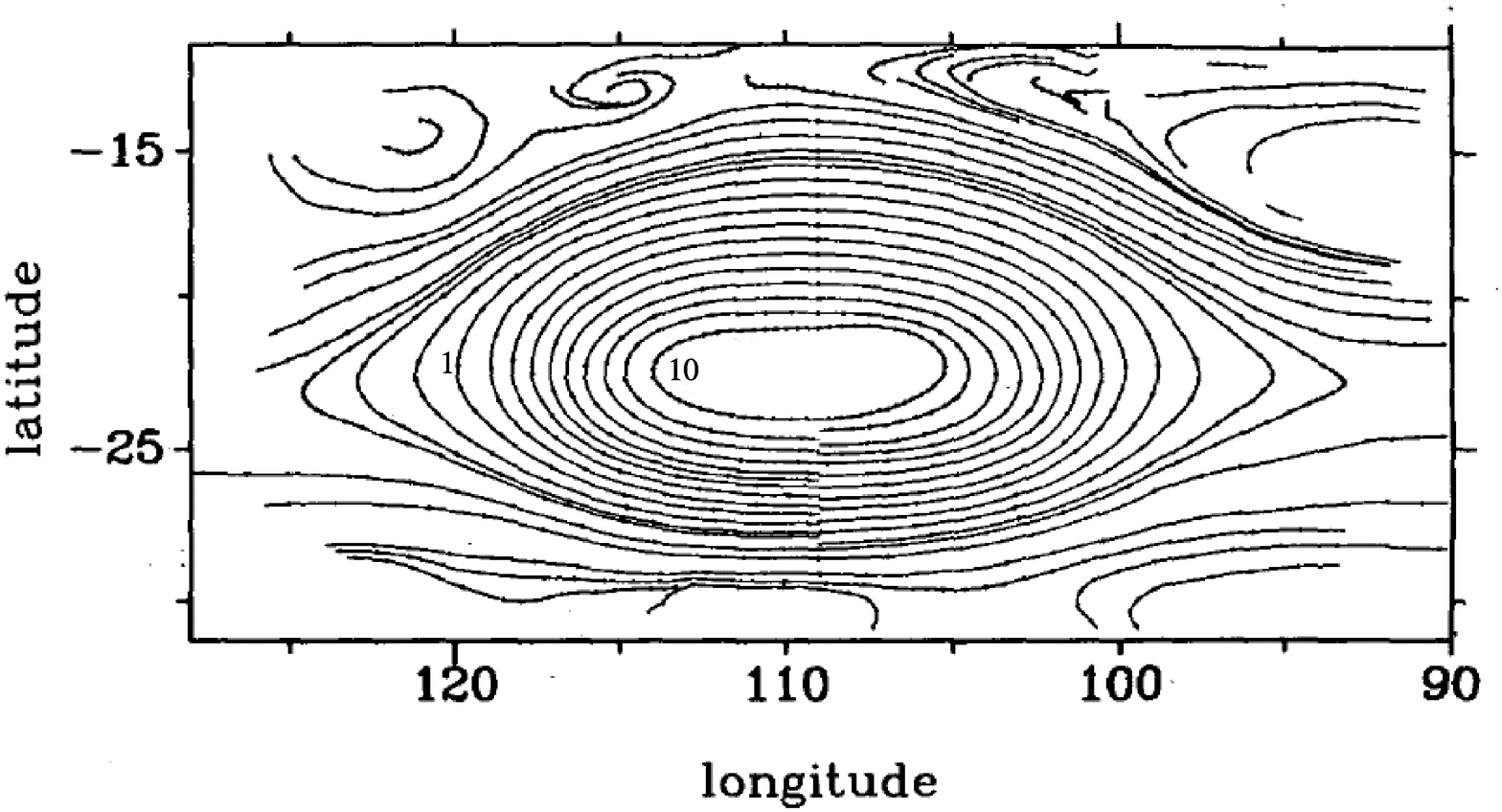}
  \caption[Dowling Figure 3]{
    \label{Figure: Dowling Fig3}
    \label{lastfig}
    Reproduction of Figure 3 from \citet{Dowling88}, showing
    trajectories calculated from \emph{Voyager} imaging data. We have
    added an annotation marking which trajectories were studied for
    this project: trajectory 1 is to the right of the numeral in the
    figure, and trajectory numbers increase sequentially with
    decreasing trajectory radius, up to trajectory 10, the inner-most
    trajectory depicted above. (For the purposes of clarity, we have
    digitally erased a trajectory that was interior to trajectory 10
    and not analyzed.) 
  }
\end{figure}

\end{document}